\documentclass[12pt]{JHEP3}
\pdfoutput=1

\usepackage{textcomp}
\usepackage{amsmath, amssymb}
\usepackage{bbold}
\usepackage{slashed}
\usepackage{graphics}
\usepackage{shadow}
\usepackage{float} 

\newcommand{\p}{P\!\!\!\!/\,\, }
\newcommand{\Ps}{\Pi\!\!\!\!/\,\, }

\title{Half-integer Higher Spin Fields in (A)dS from Spinning
  Particle Models}

\author{Olindo Corradini\\ 
Dipartimento  di Fisica, Universit{\`a} di Bologna and\\ 
INFN, Sezione di Bologna, via Irnerio 46, I-40126 Bologna, Italy\\
$\ $\&\\
Centro de Estudios en F\'isica y Matem\'aticas Basicas y Aplicadas\\
Universidad Aut\'onoma de Chiapas, Tuxtla Guti\'errez, Chiapas, Mexico\\ 

\mbox{E-mail: \email{corradini@bo.infn.it}}}

\abstract{We make use of $O(2r+1)$ spinning particle models to construct
  linearized higher-spin curvatures in (A)dS spaces for fields of
  arbitrary half-integer spin propagating in a space of arbitrary
  (even) dimension: the field potentials, whose curvatures are
  computed with the present models,
  are spinor-tensors of mixed symmetry corresponding to Young tableaux with
  $\frac D2 - 1$ rows and r columns, thus reducing to totally symmetric
  spinor-tensors in four dimensions. The paper generalizes similar
  results obtained in the context of integer spins in (A)dS.}

\preprint{}

\begin{document}

\section{Introduction}
\label{sec:intro}
Despite years of investigation higher-spin field theory
(see~\cite{review} for a review) is still not a completely understood
subject. In particular the introduction of interactions has turned
out to be a formidable task. Thus, in order to clarify such difficult
problem it may be fruitful to investigate
the subject from different perspectives such as the first-quantized
approach offered by some spinning particle models.  

Particle models constitute
an alternative to conventional (second-quantized) field theories in
the study of quantum field theory by means of the worldline
formalism~\cite{Schubert:2001he} and can also be applied in
the study of higher-spins field theory; in particular $O(N)$-extended spinning
particle models~\cite{Gershun:1979fb,Howe:1988ft} 
turn out to be related to the geometric
formulation of higher spin field theory. In the context of
second-quantized field theory, geometric equations of motions for
totally symmetric higher
spin fields were proposed in~\cite{Francia:2002aa1}
(using linearized higher spin geometry constructed
in~\cite{Weinberg:1965rz}, further developed
in~\cite{Damour:1987vm} and generalized to mixed symmetry tensors
in~\cite{DuboisViolette:2001jk}) in order to relax the algebraic 
constraints present in the conventional formulation of higher spin
equations of motion in terms of Lorentz-covariant second-order
(first-order for half integer spins) differential equation of the field
potential. For integer spins, lagrangians for totally symmetric tensors in flat space and (A)dS
space were first constructed by Fronsdal ~\cite{Fronsdal:1978rb} and later generalized to mixed
symmetry tensors in~\cite{Labastida:1986ft}, whereas  
for half-integer fields, lagrangians for totally symmetric
(spinor)-tensor fields in flat space were first proposed by Fang and
Fronsdal~\cite{Fang:1978wz} and Curtright~\cite{Curtright:1979uz}, 
whereas extensions to (A)dS spaces were soon later obtained
in~\cite{Fang:1979hq}. (Fang)-Fronsdal equations of motion are characterized
by (gamma)-trace constraints which must be satisfied by fields and
gauge parameters and can be obtained by partial gauging of the
unconstrained (compensated) geometric equations of motion.

Geometric equations of motion are a natural
outcome of spinning particles: the canonical quantization of
locally-supersymmetric $O(N)$-extended spinning particle models
\begin{eqnarray}\label{action1}
  S &=& \int dt 
  \Big [ p_\mu \dot x^\mu + \frac{i}{2} \psi_{i\mu} \dot \psi_i^\mu 
    -e \underbrace{\Big ( \frac{1}{2} p_\mu p^\mu \Big )}_{H}
    - i \chi_i \underbrace{\Big ( p_\mu \psi^\mu_i\Big )}_{Q_i} 
    - \frac{i}{2} a_{ij} \underbrace{\Big ( \psi_i^\mu \psi_{j\mu} 
      \Big )}_{J_{ij}}
    \Big ]\\ && i=1,\dots,N \nonumber  
\end{eqnarray}
 yields equations of motions for spin-$\frac N2$
fields (wave functions) in terms of the corresponding linearized
curvatures. For flat external backgrounds the correspondence is well
established (see~\cite{Marnelius:2009uw} for a recent very detailed analysis) 
and one-loop path integral results were also
obtained~\cite{Bastianelli:2007pv}. The coupling to external arbitrary
backgrounds is possible for $N\leq 2$, corresponding to worldline
descriptions of spin $\leq 1$ particles coupled to gravity~\cite{Bastianelli:2002fv}. 
 However, for $N >2$, the
coupling of the above models to external arbitrary backgrounds is
problematic~\cite{Howe:1988ft}, reflecting the aforementioned difficulty of
introducing interactions in higher spin fields. A partial way-out to
such problem was given in~\cite{Kuzenko:1995mg} where (using the
manifestly conformally-invariant
formulation of~\cite{Marnelius:1978fs,Siegel:1988ru}) it was shown how to
consistently couple $O(N)$ spinning particles to AdS backgrounds.             
In~\cite{Bastianelli:2008nm} spinning particle models with
$O(N)$-extended local supersymmetry were further analyzed and coupling
to a more generic set of backgrounds, conformally flat spaces, was
proposed: linearized higher-spin curvatures and geometric equations of
motion of integer higher-spin fields (with $N=2s,\ s\in
{\mathbb N}$) in (A)dS spaces were derived and studied, from the
canonical quantization of the models (higher-spin de Wit-Freedman linearized
curvatures for totally symmetric potentials
in AdS were constructed in~\cite{Manvelyan:2007hv}.)   

The present manuscript is a generalization of the
results obtained in~\cite{Bastianelli:2008nm} to half-integer
higher-spin fields, in (A)dS. Here we restrict our study 
to the link between $O(N)$ spinning particles and fermionic higher spin
fields. However, it is worth mentioning other particle models relevant to
higher spin physics, such as twistor-like 
superparticles~\cite{Bandos:1998vz}, $U(N)$ spinning
particles~\cite{Bastianelli:2009vj} and $Sp(2r)$ models~\cite{Bastianelli:2009eh} (obtained from
gauging orthosymplectic models~\cite{Hallowell:2007qk}) whose BRST detour
quantization~\cite{Cherney:2009mf} describes higher spins of mixed
symmetry type~\cite{Bekaert:2002dt,DuboisViolette:2001jk}. The structure of the paper is as
follows: in Section~\ref{sec:algebra} we describe the
quantum constraint algebra associated to the flat space particle
action~(\ref{action1}) and its quadratically-deformed (A)dS version
(quadratic algebras have already appeared in the study of higher spin
fields~\cite{Buchbinder:2001bs}). The geometric equations of motion
for fermionic higher spin fields in flat space and (A)dS space are
obtained in Section~\ref{sec:HSEoM} and Section~\ref{sec:HSEoM-AdS}
respectively, by imposing the constraint algebra onto the
physical wave function, hence establishing a dictionary between the
present particle formulation and the conventional second-quantized
formulation. In Section~\ref{sec:conclusions} we provide some
conclusions and future investigations that may be undertaken within
the present approach.

\section{Constraint algebra in flat space and (A)dS space}
\label{sec:algebra}
In order to obtain the quantum $O(N)$-extended supersymmetric particle
algebra (with $N=2r+1$) in flat space, from the classical constraints
of~(\ref{action1}), 
one only needs to specify the correct ordering in the definition of 
the $SO(N)$ generators $J_{ij}$, as there are no ordering ambiguities
associated with the hamiltonian $H$ and susy generators $Q_i$.
Taking that into account, the quantum constraints are given by
\begin{eqnarray}
H = \frac{1}{2} p_\mu p^\mu~,\qquad
Q_i = p_\mu \psi_i^\mu~, \qquad
J_{ij} = \frac{1}{2} [\psi_i^\mu, \psi_{j\mu}]  
\label{cons-flatsp}
\end{eqnarray}
with $\{\psi_i^\mu,\psi_j^\nu\}=\eta^{\mu\nu}\delta_{ij}$. They satisfy the quantum algebra
\begin{eqnarray}
[J_{ij},J_{kl}] &=& -\delta_{jk} J_{il} +\delta_{ik} J_{jl} 
+\delta_{jl} J_{ik} -\delta_{il} J_{jk} 
\label{3.5}
\\[1mm]
[J_{ij},Q_k] &=& \delta_{jk} Q_i - \delta_{ik} Q_j  
\label{3.4}
\\[1mm]
\{Q_i,Q_j\} &=& 2 \delta_{ij} H 
\label{3.3}
\end{eqnarray}
which is first class. The above (multi)-Clifford algebra can be
realized using gamma matrices as follows~\cite{Howe:1988ft} 
\begin{eqnarray}
  \psi_1^\mu &=& \frac1{\sqrt2}
  \gamma^\mu\otimes\gamma\cdots\otimes\gamma\,,\ \ 
\psi_2^\mu = \frac1{\sqrt2}
 {\mathbb
   I}\otimes\gamma^\mu\otimes\gamma\cdots\otimes\gamma\,,\ \ 
\psi_N^\mu = \frac1{\sqrt2}
  {\mathbb I}\cdots\otimes{\mathbb I}\otimes\gamma^\mu
\nonumber
\end{eqnarray}
with the wave function written as a multispinor
$\Psi_{\alpha_1\cdots\alpha_N}$ and with $\gamma$ being the chirality
matrix; throughout the paper we work with a generic even dimension $D=2d$. In fact
for $D=2d+1,\, N>2$ the present models are empty due to a global anomaly~\cite{Howe:1988ft}.
In $D=4$, the above constraint algebra yields Bargmann-Wigner
equations~\cite{Bargmann:1948ck} for a spin-$\frac N2$ field (BRST quantization of the
models is described in~\cite{Marnelius:1988ab}).  

Here, in order to study the spin-$s$  ($s=N/2 = r+1/2$) equations of
motion, we use a different basis taking complex combinations of  
the first $2r$ indices of $SO(N)$ and define (for $I,i=1,...,r$)
\begin{eqnarray}
  \psi^\mu_I &=& \frac{1}{\sqrt{2}}(\psi^\mu_i + i \psi^\mu_{i+r})
  \label{psiI}
  \,,\quad
  \bar \psi^\mu_{\bar I} = \frac{1}{\sqrt{2}}(\psi^\mu_i - i \psi^\mu_{i+r})
  \equiv \psi^{I\mu}
\end{eqnarray}
and 
\begin{equation}
\psi_{2r+1}^\mu \equiv\frac1{\sqrt2} \gamma^\mu~.
\label{psi-2r+1}
\end{equation}
More specifically, in order to set the antisymmetry
between~(\ref{psiI}) and~(\ref{psi-2r+1}) we define them as
\begin{eqnarray}
\psi_I^\mu &\to& \psi_I^\mu\otimes \gamma\label{psiI'}\,,\quad  
\psi^{I\mu} \to \psi^{I\mu}\otimes \gamma\,,\quad 
\gamma^\mu \to {\mathbb I}\otimes \gamma^\mu~.
\end{eqnarray}  
In the ``coordinate'' representation one can realize 
$\psi_I^\mu$ as multiplication by Grassmann variables and 
$\psi^I_\mu = \frac{\partial}{\partial \psi_I^\mu }$
(we use left derivatives). This realization keeps manifest only the $U(r)\subset SO(2r+1)$ 
subgroup of the internal symmetry group, but will be quite useful 
in classifying the constraints and their solutions. In this
representation the left space
in~(\ref{psiI'}) is the space of
antisymmetric multiforms
$\psi_1^{\mu_1}\cdots\psi_1^{\mu_{A_1}}\cdots\psi_r^{\nu_1}\cdots\psi_r^{\nu_{A_r}}$,
whereas the right space is the
fermionic space where the $(2r+1)$-th Clifford algebra is realized as gamma-matrices. In
terms of those operators we have  
\begin{eqnarray}
  \{\psi_I^\mu,\psi^{J\nu} \} &=& \eta^{\mu\nu} \delta_I^J \\
  \{\gamma^\mu,\gamma^{\nu} \} &=& 2 \eta^{\mu\nu}~. 
\end{eqnarray}
Henceforth we can avoid to explicitly write down the tensor product as its only
effect is the aforementioned antisymmetry; we will only need to take
care of imposing it.

The susy charges in the $U(r)$ basis take the form
\begin{eqnarray}
  Q_I= \psi_I^\mu p_\mu\,,\quad Q^I= \psi^{I \mu} p_\mu 
\end{eqnarray}
and  
\begin{eqnarray}
  \p=\gamma^\mu p_\mu = \sqrt2 Q_N
\end{eqnarray} 
is nothing but the Dirac operator. Hence, the non-vanishing part of the 
susy algebra~(\ref{3.3}) reads
\begin{eqnarray}
  \{Q_I, Q^J\} &=& 2 \delta_I^J H~, \qquad \p{}^2 =2H~. \label{PP}
\end{eqnarray}
In the complex basis defined above, the $SO(N)$ generators 
split as\\ $J_{ij}\sim (J_{I \bar J},J_{IJ}, J_{\bar I \bar
  J},J_{NI},J_{N\bar I})
\sim (J_I{}^J,K_{IJ}, K^{IJ}, L_I, L^I)$, which we normalize as 
\begin{eqnarray}
  &&J_I{}^J =\psi_I \cdot \psi^J -d\, \delta_I^J~, \quad
  K_{IJ} = \psi_I\cdot \psi_J~, \quad
  K^{IJ} = \psi^I \cdot \psi^J~,\quad\label{eq:so2r}
  \\&&
  L_I = \gamma^\mu\psi_{I\mu}~,\quad L^I= \gamma^\mu \psi^I_\mu\label{eq:LL}
\end{eqnarray}
so that when $I=J$, $J_I{}^J $ is a hermitian operator with real
eigenvalues; the above generators can be written in terms of the
Clifford basis by means of~(\ref{psiI}): for example
$L_I=\gamma^\mu\psi_{I\mu} = J_{N\, i}+iJ_{N\,i+r}$. The $SO(N)$
algebra (\ref{3.5}) breaks up into the $SO(2r)$ subalgebra generated
by~(\ref{eq:so2r})~\cite{Bastianelli:2008nm}.  The remaining
non-vanishing relations of the $SO(N)$ algebra, involving $L$'s are
\begin{eqnarray}
  [J_I{}^J,L_K] &=& \delta_K^JL_I~, \quad [J_I{}^J,  L^K] = -\delta^K_I
    L^J\\[2mm]
       {}[  K^{IJ},L_K] &=& \delta_K^J  L^I-\delta_K^I  L^J~,\quad
       {}[K_{IJ},  L^K] = \delta^K_J L_I-\delta^K_I L_J \label{KL}\\[2mm]
       [L_I,  L^J] &=& -2 J_I{}^J~,\quad [  L^I,  L^J] = -2   K^{IJ}~,\quad
       [L_I,L_J] = -2 K_{IJ} \label{LL}~.  
\end{eqnarray}
Finally, it is useful to list in the same basis the remaining part
of the constraint algebra corresponding to eq. (\ref{3.4})
\begin{eqnarray}
  [J_I{}^J,Q_K] &=& \delta_K^J Q_I~,\quad 
  [J_I{}^J,  Q^K]= - \delta_I^K   Q^J \\[2mm]
  {}[  K^{IJ},Q_K]&=&  \delta_K^J   Q^I  - \delta_K^I   Q^J~,\quad
  [K_{IJ},  Q^K]=  \delta_J^K Q_I  - \delta_I^K Q_J \label{KQ}\\[2mm]
  {}[  L^I,Q_J] &=& \delta_J^I \p~,\quad 
  [L_I,  Q^J]= \delta_I^J \p \label{LQ}\\[2mm]
  {}[L_I,\p] &=& -2Q_I~,\quad [  L^I,\p] = -2  Q^I~.
\end{eqnarray}
As explicitly worked out in~\cite{Bastianelli:2008nm}, the
deformation of the above flat algebra to (A)dS spaces
$R_{abcd}=b(\eta_{ac}\eta_{bd}-\eta_{ad}\eta_{bc})$, is obtained by generalizing the constraints to
\begin{eqnarray}
  &&H = \frac{1}{2}
  \left(\pi^a\pi_a-i\omega^a{}_{ab}\pi^b\right)+\frac b4
  J_{ij}J_{ij}-bA(D)~,\nonumber\\ 
  &&Q_i = \psi_i^a \pi_a = \psi_i^ae_a^\mu\left(p_\mu-\frac12 \omega_{\mu bc}M^{bc}\right)~, \qquad
  J_{ij} = \frac{1}{2} [\psi_i^a, \psi_{ja}]  
  \label{cons-adssp}
\end{eqnarray}
with $M^{ab}=\frac i2\left[\psi^a_i,\psi^b_i\right]$ being the
multispinor representation of the $SO(D)$ Lorentz generators,
$A(D)=(2-N)\frac D8-\frac{D^2}{8}$ and
\begin{eqnarray}
  \pi_\mu = p_\mu-\frac12 \omega_{\mu bc}M^{bc}\,,\quad \pi_a =e_a^\mu\pi_\mu
\end{eqnarray} 
the covariant momentum.
With these operators
equations~(\ref{3.5}-\ref{3.4}) hold unchanged, whereas the susy
algebra~(\ref{3.3}) gets modified as
\begin{eqnarray}
  \{Q_i,Q_j\} = 2\delta_{ij} H+\frac
  b2\left(J_{ik}J_{jk}+J_{jk}J_{ik}-\delta_{ij}J_{kl}J_{kl}\right)~.
  \label{eq:alg} 
\end{eqnarray}  
In the above complex basis we have $Q_I=\psi_I^a\pi_a$,
$ Q^I= \psi^{Ia}\pi_a$ and $\Ps=\gamma^a\pi_a$ 
and the susy algebra in the same basis can be obtained from~(\ref{eq:alg}) in a simple
way by noting that the complexification in the internal
indices (cfr. eq.~(\ref{psiI})) implies the transformation of the flat
metric as $\delta_{ij}\to (\delta_I{}^J,\delta^I{}_J,\delta_{NN})$. Hence, 
\begin{eqnarray}
  \{Q_I,Q_J\} &=& b\left(K_{IL} J_J{}^L+K_{JL} J_I{}^L\right)+\frac
  b4\left(L_I L_J +L_J L_I\right)\\
  \{Q_I, \Ps\} &=& b\left(K_{IK} L^K +L_KJ_I{}^K +\frac12 L_I\right)\\
  \Ps^2 &=& 2(H_0-bA(D))+\frac b2 \left( L_K L^K +  L^K L_K\right)\\
  \{ Q^I, Q^J\} &=& b\left(- K^{IL} J_L{}^J- K^{JL} J_L{}^J\right)+\frac
  b4\left( L^I  L^J + L^J  L^I\right)\\
  \{Q_I, Q^J\}&=&2\delta_I^J(H_0-bA(D))
  +\frac b4\left(L_I  L^J +  L^J L_I\right)\nonumber\\
  &&-\frac b2\left(J_I{}^K J_K{}^J+J_K{}^J J_I{}^K-K_{IK}
  K^{JK}- K^{JK} K_{IK}\right)\\
  \{ Q^I,\Ps\} &=& b\left(- L^K J_K{}^I+ K^{IK}L_K+\frac12
   L^I\right)~.
\end{eqnarray}

\section{Higher spin equations of motion in flat space}
\label{sec:HSEoM}
The equations of motion for the higher spin wave function
are obtained by imposing that the above operators (constraints) annihilate the
first-quantized physical state. A generic state in the Hilbert space
where such operators act can be written as
\begin{eqnarray}
  |R\rangle \sim a_r\sum_{A_i=0}^D\psi_1^{\mu_1}..\psi_1^{\mu_{A_1}}\cdots
  \psi_r^{\nu_1}..\psi_r^{\nu_{A_r}}
  R_{\mu_1..\mu_{A_1},\cdots,\nu_1..\nu_{A_r};\alpha}|\chi_\alpha\rangle 
\end{eqnarray} 
where $|\chi_\alpha\rangle$ is a generic state of the fermionic space,
$\gamma^\mu|\chi_\alpha\rangle = 
\gamma^\mu_{\alpha'\alpha}|\chi_{\alpha'}\rangle$, and $a_r$ is a
numerical prefactor that will be fixed shortly (cfr. eq.~(\ref{eq:ar})). In particular we
impose a minimal set of constraints on the physical state; the
remaining constraints are automatically satisfied thanks to the above
algebra. The set of constraints we impose is the following
\begin{enumerate}
\item $J_I{}^J|R\rangle =0$; for $I=J$, these constraints yield irreducibility
  conditions and pick-out from the Hilbert space the tensor  
  $R_{\mu_1..\mu_d,\cdots,\nu_1..\nu_d;\alpha}$, that is antisymmetric
  within each block of $d$ of indices and symmetric in the exchange
  of two blocks. For $I\neq J$, the constraints
  impose algebraic Bianchi identities
  $R_{[\mu_1..\mu_d,\nu_1]..\nu_d,\cdots,\lambda_1..\lambda_d;\alpha}~=~0$. 
\item $Q_I|R\rangle=0$; these constraints impose integrability
  conditions on the physical
  field. One can thus solve such conditions by writing the
  "curvature'' $|R\rangle$ in terms of a "potential''
  $|R\rangle=q|\phi\rangle$ where $q$ is a differential operator
  written in terms of the algebra operators. In terms of the
  potential the above constraints now become differential Bianchi
  identities.   
\item $ L^I |R\rangle =0$: these  constraints correspond to
  ``gamma-tracelessness'' of the higher-spin curvature
  $\gamma^\mu_{\alpha\alpha'} R_{\mu ..\mu_d,\cdots,\nu_1..\nu_d;\alpha'}=0$.  
\end{enumerate}
The remaining constraints are automatically satisfied. The trace
constraints $K^{IJ}$ are satisfied thanks to the first equation
in~(\ref{LL}); $K_{IJ}$ are satisfied because they are trace 
constraints for the dual curvature obtained using the dual basis
for the operators labelled by $I$ and $J$: it is easy to convince
oneself that such dual trace constraints are satisfied if $K^{IJ}$ are, because of the relation
$\epsilon_{\mu_1\cdots\mu_d}\epsilon^{\nu_1\cdots\nu_d}
=\delta_{\mu_1\cdots\mu_d}^{[\nu_1\cdots\nu_d]}$. In turn
 $L_I$ are satisfied thanks to the second equation in~(\ref{KL}), $\p$ is satisfied
thanks to~(\ref{LQ}) and in turn $H$ is satisfied thanks to the second
equation in~(\ref{PP}). Finally $ Q^I$ are satisfied thanks to the
first equation in~(\ref{KQ}).~\footnote{For spin 3/2 ($r=1$) there are no traces. In
  this case it is easy to prove that
  $\gamma^{\mu_1}R_{\mu_1\cdots\mu_d}=0\Rightarrow
  \gamma\gamma^{\nu_2\cdots\nu_d\mu_1\cdots\mu_d} 
  R_{\mu_1\cdots\mu_d} =\epsilon^{\nu_1\cdots\nu_d\mu_1\cdots\mu_d}
  \gamma_{\nu_1}R_{\mu_1\cdots\mu_d}=0$. The first term and the last
  term are respectively $L^1|R\rangle =0$ and  $L_1|R\rangle=0$
  expressed in components.} 

For $D$ and $N$
  arbitrary, the above constraints correspond to the free equations of
  motion of a conformal particle in flat $D$
  dimensions~\cite{Siegel:1988ru}: the algebraic constraints pick out
  a representation of the conformal group $SO(2,D)$ represented by the
  rectangular Young tableau with $d=\frac D2$ rows and $r$ columns.

\subsection*{Higher-spin curvature}
In flat space it is easy to solve the integrability conditions since
$Q_I$'s anticommute. Hence
\begin{eqnarray}
  |R\rangle = q |\phi\rangle~,\quad q =\frac1{r!}\epsilon^{I_1\cdots I_r}
  Q_{I_1}\cdots Q_{I_r}~,
\end{eqnarray}
with $J_I{}^J|\phi\rangle = -\delta_I^J|\phi\rangle$ and 
\begin{eqnarray}
  |\phi\rangle\sim
  \psi_1^{\mu_1}..\psi_1^{\mu_{d-1}}\cdots
  \psi_r^{\nu_1}..\psi_r^{\nu_{d-1}}~ 
  \phi_{\mu_1..\mu_{d-1},\cdots,\nu_1..\nu_{d-1};\alpha}|\chi_\alpha\rangle~. 
\end{eqnarray}
In components it reads
\begin{eqnarray}
  R_{\mu_1..\mu_d,\cdots,\nu_1..\nu_d;\alpha}
  &=&\partial_{\mu_1}\cdots\partial_{\nu_1}\phi_{\mu_2..\mu_d,\cdots,\nu_2..\nu_d;\alpha}
\end{eqnarray}
with implied antisymmetrization among each block of indices and
symmetrization between exchange of two blocks. Above we fixed
\begin{eqnarray}
  a_r=\left\{\begin{array}{ll} 1\,, &\quad r\ {\rm even}\\ -i\,, &\quad
  r\ {\rm odd}\end{array}\right.
  \label{eq:ar}
\end{eqnarray}
in order to obtain a real-valued normalization for the curvature.

\subsection*{Higher-derivative equations of motion and their gauge invariance}
The gamma-trace conditions imposed upon the curvature yield the
higher-curvature equations of motion satisfied by the associated gauge
potential. By rewriting the curvature operator as 
\begin{eqnarray}
  q=\frac1r Q_J  q^J=Q_1  q^1~,\quad  q^I \equiv \frac1{(r-1)!}\epsilon^{II_2\cdots
    I_r}Q_{I_2}\cdots Q_{I_r}
  \label{eq:barq}
\end{eqnarray}
the gamma-trace constraints read
\begin{eqnarray}
   L^1 q|\phi\rangle =  L^1 Q_1  q^1 |\phi\rangle =0
\end{eqnarray}
and, using the above algebra, it reduces to
\begin{eqnarray}
  (-)^{r-1} q^1\Big(\p+Q_1 L^1\Big)|\phi\rangle =
  (-)^{r-1} q^1\Big(\p+Q_K L^K\Big)|\phi\rangle =0 
\end{eqnarray}
where in the last equality we used the nilpotency condition of the $Q_I$'s, so that
\begin{eqnarray}
  q^I {\cal F}|\phi\rangle =  L^I q|\phi\rangle =0
\label{eq:HDEoM}
\end{eqnarray}
where 
\begin{eqnarray}
  {\cal F} = (-)^{r-1}\Big(\p +Q_K L^K\Big)
\end{eqnarray}
is the Fang-Fronsdal operator~\cite{Fang:1978wz} that is the
higher-spin generalization of the Dirac and Rarita-Schwinger operators and the expression  
\begin{eqnarray}
   L^I|R\rangle =  q^I{\cal F} |\phi\rangle
\end{eqnarray}
is the half-integer higher-spin generalization of the Damour-Deser
identity~\cite{Damour:1987vm}. 

By defining the gauge transformation
\begin{eqnarray}
  \delta_\xi|\phi\rangle = Q_K  V^K|\xi\rangle 
  \label{eq:gauge-transf-f}
\end{eqnarray}
where $ V^K = V^\mu(x)\psi^K_\mu$ and $|\xi\rangle$ being a
tensor of the same species as $|\phi\rangle$, we have
\begin{eqnarray}
  \delta_\xi{\cal F}|\phi\rangle = (-)^{r-1}Q_I Q_J  V^J  L^I |\xi\rangle
  \label{gauge-transf-FF}
\end{eqnarray}
that is not trivial provided $r\geq 2$ ($s\geq
5/2$). From~(\ref{gauge-transf-FF}) the gauge invariance of the
higher-derivative equations of motion~(\ref{eq:HDEoM}) immediately
follows
\begin{eqnarray}
  \delta_\xi\,  q^I {\cal F}|\phi\rangle = 0
\end{eqnarray} 
thanks again to the nilpotency of $Q_I$'s. Such a property also yields
\begin{eqnarray}
  {\cal F}|\phi\rangle = Q_I Q_J  W^J  W^I |\rho\rangle
\label{eq:EoMC}
\end{eqnarray}
as generic kernel of $q^I$ in~(\ref{eq:HDEoM}). The latter is
gauge-invariant provided the ``compensator'' field transforms as 
\begin{eqnarray}
  \delta_\xi  W^J  W^I|\rho\rangle = (-)^{r-1} V^{[J}  L^{I]}|\xi\rangle~.
  \label{eq:comp-gauge}
\end{eqnarray}

\subsection*{Fang-Fronsdal equation of motion: partial gauge fixing}
The Fang-Fronsdal linear equation of motion 
\begin{eqnarray}
  {\cal F} |\phi\rangle =0
\end{eqnarray}
for a free spin-$s$ field can be obtained from the compensated linear equation~(\ref{eq:EoMC})
by gauging away the compensator $ V^{[J} 
  L^{I]}|\xi\rangle=- W^J  W^I|\rho\rangle$. This condition,
and in turn the Fang-Fronsdal equation, are
preserved by a gauge symmetry, given by~(\ref{eq:gauge-transf-f})
subject to the algebraic constraint  
\begin{eqnarray}
  Q_J Q_K  V^K  L^J |\xi\rangle =0 
\end{eqnarray}
that is non-trivial when $r\geq 2$ ($s\geq 5/2$) and  
corresponds to gamma-tracelessness of the gauge parameter
\begin{eqnarray}
   L^I|\xi\rangle =0~. 
\end{eqnarray}
In turn, the gauge potential must satisfy algebraic constraints. After
a little algebra one obtains 
\begin{eqnarray}
  0= \Biggl(Q_I L^I\p + \frac23 Q_I Q_J  L^I  L^J\Biggr) {\cal
    F}|\phi\rangle = (-)^r Q_I Q_J Q_K  L^I  L^J  L^K |\phi\rangle 
\label{eq:triplegamma}
\end{eqnarray}
that is non-trivial when $s\geq 7/2$ and corresponds to triple gamma-tracelessness of the gauge
potential.   

We thus see how neatly the present complex basis of the
$SO(N)$-extended spinning particle constraint algebra reproduces all
the known features of free higher spin field theory in flat space.

\subsection*{An example: spin-$\frac72$ in four dimensions}
To make contact with the standard notation we specialize to the lowest
spin that develops all the features described above,
spin-$\frac72$. In such a case the curvature is given by
\begin{eqnarray}
  |R\rangle = Q_1Q_2Q_3 |\phi\rangle
  \quad\to\quad R_{\lambda_1\lambda_2\ \mu_1\mu_2\ \nu_1\nu_2\,;\alpha}
  =\partial_{\lambda_1}\partial_{\mu_1}\partial_{\nu_1} \phi_{\lambda_2\,\mu_2\,\nu_2\,;\alpha}
\end{eqnarray}  
with implied antisymmetrization within pairs of indices
(e.g. $\lambda_1\leftrightarrow \lambda_2$) and symmetry between
exchange of pairs (e.g. $\lambda_1\lambda_2 \leftrightarrow
\mu_1\mu_2$).  The cubic equation of motion following from
gamma-tracelessness is 
\begin{eqnarray}
  && L^1 |R\rangle = q^1{\cal F}|\phi\rangle =0 \nonumber\\ &&\to\ 
  \partial_{\mu_1}\partial_{\nu_1}\Big(\partial\!\!\!/
  \phi_{\lambda_2\,\mu_2\,\nu_2}-\partial_{\lambda_2}\gamma^\lambda 
  \phi_{\lambda\,\mu_2\,\nu_2} -\partial_{\mu_2}\gamma^\mu 
  \phi_{\lambda_2\,\mu\,\nu_2}-\partial_{\nu_2}\gamma^\nu 
  \phi_{\lambda_2\,\mu_2\,\nu}\Big)=0
 \end{eqnarray}  
where spinorial indices have been suppressed and $\phi$ is a totally
symmetric tensor; here Damour-Deser
identity and Fang-Fronsdal linear operator are self-evident. The latter
equation is gauge invariant under the unconstrained gauge symmetry
\begin{eqnarray}
  \delta |\phi\rangle = Q_K V^K |\xi\rangle \quad\to\quad
  \delta \phi_{\lambda_1\,\mu_1\,\nu_1} 
  =\partial_{\lambda_1}\zeta_{\mu_1\,\nu_1}+\partial_{\mu_1}\zeta_{\lambda_1\,\nu_1}
  +\partial_{\nu_1}\zeta_{\lambda_1\,\mu_1}  
\end{eqnarray}
thanks to the above antisymmetry; here $\zeta_{\mu\,\nu} = -iV^\lambda \xi_{\lambda\,\mu\,\nu}$.
The cubic equation of motion can be reduced to a compensated linear
equation
\begin{eqnarray}
  {\cal F}|\phi\rangle &=& Q_IQ_J  W^J  W^I|\rho\rangle \nonumber\\
  & \downarrow&\\
  \partial\!\!\!/\phi_{\lambda\,\mu\,\nu}&-&\partial_{\lambda}\gamma^{\lambda'} 
  \phi_{\lambda'\,\mu\,\nu} -\partial_{\mu}\gamma^{\mu'} 
  \phi_{\lambda\,\mu'\,\nu}-\partial_{\nu}\gamma^{\nu'} 
  \phi_{\lambda\,\mu\,\nu'}
  =-2\Big(\partial_\lambda\partial_\mu\sigma_\nu+\partial_\nu\partial_\lambda\sigma_\mu
  +\partial_\mu\partial_\nu\sigma_\lambda\Big)\nonumber 
\end{eqnarray}
with $\sigma_\nu = W^\lambda W^\mu \rho_{\lambda\mu\nu}$ being the
compensator field. The latter is gauge-invariant with unconstrained
parameter if the compensator field transforms as $\delta \sigma_\nu =
\gamma^{\mu'}\zeta_{\mu'\nu}$. Hence the Fang-Fronsdal equation 
\begin{eqnarray}
\partial\!\!\!/\phi_{\lambda\,\mu\,\nu}&-&\partial_{\lambda}\gamma^{\lambda'} 
  \phi_{\lambda'\,\mu\,\nu} -\partial_{\mu}\gamma^{\mu'} 
  \phi_{\lambda\,\mu'\,\nu}-\partial_{\nu}\gamma^{\nu'} 
  \phi_{\lambda\,\mu\,\nu'} =0
\end{eqnarray}
is consistent provided $\gamma^\lambda \gamma^\mu \gamma^\nu
\phi_{\lambda\mu\nu} =0$ (cfr. eq.~(\ref{eq:triplegamma})) 
and it is gauge invariant provided $\gamma^{\mu'}\zeta_{\mu'\nu}=0$.

\section{Higher spin equations of motion in (A)dS space}
\label{sec:HSEoM-AdS}
Similarly to the flat space case discussed in the previous section, 
geometric equations for higher
spins in (A)dS can be obtained by imposing the $SO(2r+1)$ spinning algebra generators
as constraints on the wave function. In the present case, as discussed in
Section~\ref{sec:algebra}, the algebra is a quadratic deformation of
the flat Lie algebra: it is no more a Lie algebra but is still first
class. In order to better solve such constraints, we found it convenient to
``rotate'' the $SO(2r)$ susy generators
\begin{eqnarray}
  {\cal Q}^{(\pm)}_{I}= Q_I\pm\frac{\sqrt b}2 L_I
\end{eqnarray}
that satisfy the anti-commutation relations
\begin{eqnarray}
  {\cal Q}^{(+)}_{I}{\cal Q}^{(-)}_{J}+{\cal Q}^{(+)}_{J}{\cal
    Q}^{(-)}_{I}= b\left(K_{IL}J_J{}^L+K_{JL}J_I{}^L\right) =  {\cal
    Q}^{(-)}_{I}{\cal Q}^{(+)}_{J}+{\cal Q}^{(-)}_{J}{\cal Q}^{(+)}_{I}~. 
  \label{eq:q+q-}
\end{eqnarray}
and have the same commutation properties with the $SO(2r)$ generators
$(J_I{}^J$, $K_{IJ},$ $ K^{IJ})$ as had the original susy
operators $Q_I$.
 
The set of independent constraints we now impose is:
\begin{enumerate}
\item $J_I{}^J|R\rangle =0$; irreducibility
  conditions $+$ algebraic Bianchi identities: it selects the same
  $GL(D)$ Young tableau as in flat space.
\item ${\cal Q}^{(-)}_I|R\rangle=0$; integrability
  conditions: it yields a gauge-invariant curvature. 
\item $ L^I |R\rangle =0$; gamma-tracelessness: it yields
  higher-curvature equation of motion and Damour-Deser identity.
\end{enumerate}
All other constraints are satisfied thanks to the (A)dS-deformed
spinning particle algebra described in Section~\ref{sec:algebra}.

\subsection*{Higher-spin curvature}
The explicit expression for the linearized higher-spin curvature in
(A)dS, a polynomial in the (A)dS scale $b$,  can be
obtained by solving the above integrability condition 
\begin{eqnarray}
  {\cal Q}^{(-)}_I|R\rangle=0 
  \label{eq:Bianchi-AdS}
\end{eqnarray}
  and reads 
\begin{eqnarray}
   |R\rangle =
  \sum_{n=0}^{[r/2]}(-b)^n r_n(r) q_n(r)|\phi\rangle  
  \label{eq:curv-AdS}
\end{eqnarray}
where the operators $q_n(r)$ are
now given by (recall that $r=s-1/2=[s]$)
\begin{eqnarray}
  q_n(r) = \frac1{r!}\epsilon^{I_1\cdots I_r}\times\left\{
  \begin{array}{ll}
    K_{I_1 I_2}\cdots K_{I_{2n-1}I_{2n}}~{\cal
      Q}^{(+)}_{I_{2n+1}}{\cal Q}^{(-)}_{I_{2n+2}}\cdots{\cal Q}^{(-)}_{I_{r}}~,\quad & r=2p\\[3mm] 
    K_{I_1 I_2}\cdots K_{I_{2n-1}I_{2n}} ~{\cal
      Q}^{(+)}_{I_{2n+1}}{\cal Q}^{(-)}_{I_{2n+2}}\cdots{\cal Q}^{(+)}_{I_{r}}~,\quad & r=2p+1 
  \end{array}\right.
  \label{eqqn}
\end{eqnarray}  
and with $r_n(r)$ numerical coefficients recursively
given in terms of the Pochhammer function 
$P(r,k)\equiv r(r-1)(r-2) \cdots (r-k)$ as follows
\begin{eqnarray}
  r_n(r) = \frac1{2n}\sum_{k=1}^n r_{n-k}(r)\, a_{2k}(r-2(n-k)+1)~,  
  \qquad r_0(r)\equiv 1
  \label{rn-1}
\end{eqnarray} 
where
\begin{eqnarray}
  a_{2k}(r) = f_k~P(r,2k)= f_k~\prod_{l=0}^{2k}(r-l)
  \label{eq:a2k}
\end{eqnarray}
and the $r$-independent coefficients $f_k$ are generated by
the Taylor expansion of the tangent function, 
$\tan(z)=\sum_{k=0}^\infty f_k~z^{2k+1} $.

In the integer spin case~\cite{Bastianelli:2008nm} the
anti-commutation relations of susy generators $Q_I$ were precisely the same
as~(\ref{eq:q+q-}) and, along with
the irreducibility conditions $J_I{}^J|R\rangle=0$, were the
only needed relations to solve the integrability conditions in
(A)dS and obtain the higher spin curvatures. 
Hence, due to the anti-commutation relations~(\ref{eq:q+q-})
the above operators $q_n(r)$ satisfy the same identities  as did their integer spin
counterparts (see Appendix
B of~\cite{Bastianelli:2008nm}). Therefore
linear combinations of operators $q_n(r)$ with the {\em same}
coefficients $r_n(r)$ as in the integer spin case (with the
replacement $s\to r$) must satisfy the above Bianchi identity.   
In summary, the correspondence between the integer case and the present
half-integer case amounts to the replacements

$$
\begin{array}{l | c | c}\hline &&\\
  {\rm Spin} & s\in {\mathbb N} &\ s=r+1/2,\ r\in {\mathbb
    N}\\&&\\
\hline &&\\
  {\rm Operators}\ q_n &\ \frac1{r!}\epsilon^{I_1\cdots
    I_s}K_{I_1I_2}\cdots K_{I_{2n-1}I_{2n}} Q_{I_{2n+1}}\cdots
  Q_{I_s}\ &
{\rm eq.}~(4.5)
\\
&&\\\hline &&\\
  {\rm Coefficients} & r_n(s) & r_n(r)\\&&\\
\hline &&\\
  {\rm Bianchi\ identity}\ & Q_I|R\rangle =0 & {\cal
    Q}_I^{(-)}|R\rangle=0\\&&\\
\hline &&\\
  {\rm Gauge\ transf.'s}\ & \delta_\xi|\phi\rangle =Q_K
  V^K|\xi\rangle & {\rm eq.}~(4.9)\\&&\\
\hline
\end{array}
$$

\subsection*{Equations of motion and their gauge invariance}
In the previous section we solved the curvatures in terms of their potential, thus
explicitly implementing differential Bianchi identities. Now we can impose  
gamma traces upon the curvatures to get higher-derivative equations of motion
\begin{eqnarray}
   L^I |R\rangle = \sum_{n=0}^{[r/2]}(-b)^n r_n(r)  L^I q_n(r)|\phi\rangle=0
\end{eqnarray} 
Gauge transformations of the gauge potentials are given by
\begin{eqnarray}
  \delta_\xi|\phi\rangle =\left\{
  \begin{array}{ll}
    {\cal Q}_I^{(+)}  V^I|\xi\rangle~,&\quad r=2p\\[3mm]
    {\cal Q}_I^{(-)}  V^I|\xi\rangle~,&\quad r=2p+1
  \end{array}\right.~. 
  \label{eq:gauge-transf-AdS}
\end{eqnarray}
The latter difference may appear bizarre but it is simply due to the
fact that the number of $\psi$'s in the two cases above differ by
one. Hence, if we write the transformation in components,  
when the $\psi$'s pass to the left of the gamma matrix the sign difference
cancels, giving rise to the same gauge transformation as given in~\cite{Fang:1978wz}.    

Once again the gamma-trace constraint yields the equation of motion
for the field potential
\begin{eqnarray}
   L^I |R\rangle =  q^I {\cal F}_r|\phi\rangle=0
  \label{eq:DD-AdS}
\end{eqnarray}
where $ q^I$ is the (A)dS counterpart of the operator defined
in~(\ref{eq:barq}) and the ($D$-dimensional generalization of the) 
Fang-Fronsdal differential operator is expected to be
\begin{eqnarray}
  {\cal F}_r &=& (-)^{r-1}\Big(\Ps+Q_K L^K\Big)+\frac{\sqrt b}{2}
  L^K L_K\nonumber\\ &=& (-)^{r-1}\Big(\Ps+{\cal
    Q}^{(\varepsilon_r)}_K L^K\Big) +\sqrt b J_K{}^K~,
  \label{eq:Fr-AdS} 
\end{eqnarray} 
with $\varepsilon_r =(-)^{r-1} $. In fact we note that,
using~(\ref{eq:gauge-transf-AdS}), the gauge transformation of ${\cal
  F}_r|\phi\rangle$ reads
\begin{eqnarray}
  \delta_\xi {\cal F}_r|\phi\rangle = (-)^{r-1} \Big( {\cal
    Q}_{[I}^{(-)}{\cal Q}_{J]}^{(+)}-bK_{IJ}\Big)  V^J  L^I
  |\xi\rangle
  \label{eq:FF-AdS-gauge}
\end{eqnarray}
and is $D$-independent. Morevoer, for $r\geq 2~(s\geq 5/2)$, it 
vanishes only if the gauge parameter is gamma-traceless, $ L^I |\xi\rangle =0$.

\subsection*{Examples}
In the remainder of the section, in order to test our results, 
we explicitly prove the gauge invariance of the
higher spin curvatures and obtain the Damour-Deser identities for the
simplest cases,\\ $r=1,2$.

\subsubsection*{(i) Spin $3/2$}
With $r=1$ we simply have   
\begin{eqnarray}
  |R\rangle = q|\phi\rangle= {\cal Q}^{(+)}|\phi\rangle 
  \label{eq:spin32-curv-ads}
\end{eqnarray}
so that Bianchi identity and gauge invariance
\begin{eqnarray}
  {\cal Q}^{(-)}q|\phi\rangle=q{\cal Q}^{(-)} V|\xi\rangle
  =0~,\quad \forall \xi 
  \label{eq:spin32-bianchi-gauge}
\end{eqnarray}
are obvious thanks to the identities ${\cal Q}^{(-)}{\cal Q}^{(+)}={\cal
  Q}^{(+)}{\cal Q}^{(-)} =0$. The Damour-Deser identity in this case
is trivial
\begin{eqnarray}
   L |R\rangle = {\cal F}_1|\phi\rangle= 0
  \label{eq:spin32-FF-ads}
\end{eqnarray}
as the gamma-trace of the curvature is linear in derivatives and it is
thus identically equal to the Fang-Fronsdal equation of motion.

In order to make contact with conventional notation 
let us re-write the previous expressions in components: we just need to pull
out the fermionic coordinates $\prod \psi^\mu$. 
Equations~(\ref{eq:spin32-curv-ads}-\ref{eq:spin32-bianchi-gauge}) read
\begin{eqnarray}
  && R_{\mu_1\cdots\mu_d} =
  \nabla_{[\mu_1}\phi_{\mu_2\cdots\mu_d]}\\
  && \nabla_{[\mu_1} \nabla_{\mu_2}
    \phi_{\mu_3\cdots\mu_{d+1}]} = \nabla_{[\mu_1} \nabla_{\mu_2}
    \zeta_{\mu_3\cdots\mu_d]}=0 \label{eq:spin32-B-comp}
\end{eqnarray}
where $ \nabla_\mu =D_\mu+ i(-)^d\frac{\sqrt b}{2}\gamma_\mu $
is the so-called ($SO(2,D-1)$) $SO(1,D)$ covariant derivative in (A)dS
($D_\mu$ is the
standard covariant derivative in (A)dS) and
$\zeta_{\mu_1\cdots\mu_{d-2}} = -iV^{\mu}\,\xi_{\mu\mu_1\cdots\mu_{d-2}}$. The commutator of two
$\nabla$'s acts trivially on the spinor index whereas it acts as
the standard commutator on space-time indices. Hence, differential
Bianchi identity and gauge invariance
(cfr. eq.~(\ref{eq:spin32-B-comp})) are guaranteed thanks to standard
algebraic Bianchi identity obeyed by the (A)dS Riemann tensor. The
gauge-invariant Fang-Fronsdal equation~(\ref{eq:spin32-FF-ads}) in components reads
\begin{eqnarray}
D\!\!\!\!/~
\phi_{\mu_2\cdots\mu_d}+i(-)^d\sqrt{b}~\phi_{\mu_2\cdots\mu_d}-(d-1)\tilde\nabla_{[\mu_2}
  \gamma\cdot\phi_{\mu_3\cdots\mu_d]} =0
\end{eqnarray}
with $\tilde\nabla_\mu =D_\mu- i(-)^d\frac{\sqrt b}{2}\gamma_\mu $.
 In four dimensions, $d=2$, the previous equations reduce to the known
gravitino equations in (A)dS.  

\subsubsection*{(ii) Spin $5/2$}
This case is the simplest case where most features appear
non-trivially. For the curvature, from equations~(\ref{eq:curv-AdS}-\ref{eq:a2k}), one
explicitly obtains 
\begin{eqnarray}
  |R\rangle = q|\phi\rangle= \frac12\epsilon^{I_1I_2}\Biggl({\cal
    Q}^{(+)}_{I_1} {\cal Q}^{(-)}_{I_2}-bK_{I_1I_2}\Biggr)|\phi\rangle 
  \label{eq:curv-r2}
\end{eqnarray}
whose gamma-traceless conditions yield
 \begin{eqnarray}
   L^I|R\rangle =  q^I {\cal F}_2 |\phi\rangle =0~,
  \label{eq:DD-r2}
\end{eqnarray}
with $ q^I =\epsilon^{IJ}{\cal Q}_J^{(+)}$. Above the first equality
is the Damour-Deser identity  and the second 
is the higher-curvature equation of motion for the spin-$5/2$ in (A)dS.
The expression~(\ref{eq:curv-r2}) is by construction Bianchi-identical with respect to
${\cal Q}_I^{(-)}$, and it not difficult to check
that~(\ref{eq:curv-r2}) and~(\ref{eq:DD-r2}) are gauge invariant with
respect to~(\ref{eq:gauge-transf-AdS}), with unconstrained parameter.  
In turn, this yields the unconstrained compensated equation of motion 
\begin{eqnarray}
  {\cal F}_2|\phi\rangle = \Big( {\cal
    Q}_I^{(-)}{\cal Q}_J^{(+)}-bK_{IJ}\Big)  W^J  W^I |\rho\rangle
  \label{eq:EoMconstr-AdS}
\end{eqnarray}
as kernel of $ q^I$, that is gauge-invariant provided
\begin{eqnarray}
  \delta_\xi  W^J  W^I |\rho\rangle= - V^{[J}  L^{I]} |\xi\rangle
  \label{eq:transf-constr-AdS}
\end{eqnarray}
that has the same form as its flat counterpart~(\ref{eq:comp-gauge}); 
the Fang-Fronsdal
equation of motion ${\cal F}_2|\phi\rangle=0$, obtained by gauging
away the compensator, is preserved by gamma-traceless gauge
transformations.

\subsubsection*{(iii) Spin $> 5/2$}
The explicit forms of the higher spin curvatures are again given
by~(\ref{eq:curv-AdS}-\ref{eq:a2k}). Although for these cases we do
not have an explicit form of the operator $ q^I$ appearing in the Damour-Deser
identity~(\ref{eq:DD-AdS}), we do know the Fang-Fronsdal
operators in (A)dS spaces and their gauge transformations
(cfr. eq.'s~(\ref{eq:Fr-AdS}-\ref{eq:FF-AdS-gauge})); 
we can thus infer that equations~(\ref{eq:EoMconstr-AdS}) and~(\ref{eq:comp-gauge})
hold unchanged for generic spin in (A)dS.

\section{Conclusions and Outlook}
\label{sec:conclusions}
In the present manuscript (a generalization
of~\cite{Bastianelli:2008nm}) we used locally supersymmetric $O(2r+1)$-extended
spinning particle models in maximally symmetric $D$-dimensional spaces to compute
higher spin linearized curvatures for some half-integer spin fields
(those corresponding to Young tableaux with $\frac D2 -1$ rows and $r$ 
columns) and we used them to analyze higher spin equations of motion
for some specific cases. Equations~(\ref{eq:curv-AdS}-\ref{eq:a2k}) are the main results
of the manuscript. 

The $O(N)$-extended spinning particles are easily seen to be Weyl
invariant for all~$N$: using this property  
in~\cite{Bastianelli:2008nm} it was shown how they consistently
propagate in generic conformally flat spaces. In fact the associated constraint
algebra keeps being first class, though in a non-linear way. This
might indicate that the present analysis might have an extension to
conformally flat spaces. However, although 
BRST lagrangian constructions for propagation of (massive) spin-3/2 field and spin-2 in spaces
more generic than maximally symmetric ones were recently
constructed~\cite{Buchbinder:2010gp}, for spin larger than two, the
higher spin counterparts of such BRST algebras appear to close only
for maximally symmetric spaces. So, even if one were able to solve the
above conformally flat spinning particle algebra in terms of a
higher-derivative equation of motion for the higher spin potential, it
might be problematic to cast the solution into a compensated
first-order (second-order for integer spins) equation of motion: it
would be very interesting to further clarify this point.

Finally, it would also be quite interesting to find connections
between the present particle models and the geometric approach to massive higher-spin
theories~\cite{Francia:2007ee}; in particular, by dimensionally reducing the results
obtained in~\cite{Bastianelli:2008nm} and here,  
it seems natural to link $O(N)$-extended particle models to
massive higher-spin theories~\cite{Deser:2001pe}
defined in odd-dimensional spaces~\cite{Plyushchay:1991qd}.    

\acknowledgments{This work was partly supported by the Italian
  MIUR-PRIN contract 20075ATT78. The author would like to thank
  F. Bastianelli for help, discussions and careful reading of
  the manuscript and E. Latini for discussions.}


\end{document}